\documentclass[fleqn,10pt]{article}
\usepackage{amsmath,amssymb,amsfonts}

\usepackage{amsfonts,amssymb,cite}

\newcommand{\bls}[1]{\renewcommand{\baselinestretch}{#1}}

\def\noi{\noindent}

\newcommand{\Title}[1]{\noi {{\Large\bf #1}}\\[1ex]}

\newcommand{\Author}[2]{\noi{\bf #1}\\[2ex]\noi{\normalsize\it #2}\\}

\newcommand{\Abstract}[1]{\vskip 2mm \begin{center}
        \parbox{16.4cm}{\small\noi #1} \end{center}\medskip}

\def\nqq{\hspace*{-2em}}

\def\cm{\hspace*{1cm}}

\def\Jl#1#2{#1 {\bf #2},\ }

\def\ApJ#1 {\Jl{Astroph. J.}{#1}}
\def\CQG#1 {\Jl{Class. Quantum Grav.}{#1}}
\def\DAN#1 {\Jl{Dokl. AN SSSR}{#1}}
\def\GC#1 {\Jl{Grav. Cosmol.}{#1}}
\def\GRG#1 {\Jl{Gen. Rel. Grav.}{#1}}
\def\JETF#1 {\Jl{Zh. Eksp. Teor. Fiz.}{#1}}
\def\JETP#1 {\Jl{Sov. Phys. JETP}{#1}}
\def\JHEP#1 {\Jl{JHEP}{#1}}
\def\JMP#1 {\Jl{J. Math. Phys.}{#1}}
\def\NPB#1 {\Jl{Nucl. Phys. B}{#1}}
\def\NP#1 {\Jl{Nucl. Phys.}{#1}}
\def\PLA#1 {\Jl{Phys. Lett. A}{#1}}
\def\PLB#1 {\Jl{Phys. Lett. B}{#1}}
\def\PRD#1 {\Jl{Phys. Rev. D}{#1}}
\def\PRL#1 {\Jl{Phys. Rev. Lett.}{#1}}

\def\lal{&&\nqq {}}
\def\eq{Eq.\,}
\def\eqs{Eqs.\,}
\def\beq{\begin{equation}}
\def\eeq{\end{equation}}
\def\bear{\begin{eqnarray}}
\def\bearr{\begin{eqnarray} \lal}
\def\ear{\end{eqnarray}}
\def\earn{\nonumber \end{eqnarray}}

\def\yy{\\[5pt] {}}
\def\yyy{\\[5pt] \lal }

\def\dst{\displaystyle}

\def\fracd#1#2{{\dst\frac{#1}{#2}}}

\def\Half{{\fracd{1}{2}}}

\def\diag{\mathop{\rm diag}\nolimits}

\def\const{{\rm const}}

\def\rf{\eqref}
\def\mn{_{\mu\nu}}
\def\MN{^{\mu\nu}}
\def\mN{_\mu^\nu}

\def\kappa{\varkappa}

\def\GR{general relativity}

\def\ssph{static, spherically symmetric}

\def\bh{black hole}

\bls{1.1}
\begin{document}
\thispagestyle{empty}
\twocolumn[


\Title{Comment on ``Linear superposition of regular black hole solutions\yy
		of Einstein nonlinear electrodynamics''}
	
\Author{K. A. Bronnikov}
	{VNIIMS, Ozyornaya ul. 46, Moscow 119361, Russia;\\
	 Inst. of Gravitation and Cosmology, RUDN University,
         ul. Miklukho-Maklaya 6, Moscow 117198, Russia;\\
	National Research Nuclear University ``MEPhI'', Kashirskoe sh. 31, Moscow 115409, Russia}

\Abstract
   {It is argued that in the paper by A. A. Garcia-Diaz and G. Gutierrez-Cano
   [\PRD {100} 064068 (2019)]  on nonlinear electrodynamics coupled to general relativity, 
   along with some interesting results and useful observations, many statements are either 
   inaccurate or incomplete. In particular, the authors only consider solutions with an
   electric charge, whereas their magnetic counterparts have features of equal interest,
   both similar to and different from those of electric ones. Moreover, it is not mentioned that in 
   electric solutions with a regular center the Lagrangian function $L(f)$ ($f = F\mn F\MN$) cannot 
   have a Maxwell weak-field limit. The observation on superpositions of regular solutions suffers
   some inaccuracies. The present Comment tries to fill these and other gaps and to provide necessary 
   corrections.
   }
\medskip 
] 

  In their paper \cite{GDC}, A.A. Garcia-Diaz and G. Gutierrez-Cano focused on the properties of 
  static, spherically, planarly and pseudospherically symmetric metrics of \GR\ (GR) coupled to
  nonlinear electrodynamics (NED) that, in their opinion, had been previously unnoticed. In particular: 
\begin{enumerate}  \itemsep -1pt
\item Extension of the Birkhoff theorem to Einstein-NED space-times. 
\item Determination of the algebraic types of the NED stress-energy tensor (SET) 
       $T\mN$, hence of the Einstein tensor $G\mN$. 
\item A formulation of the inverse integration method.
\item The existence of linear superpositions of Einstein-NED solutions with given metric functions. 
\item A generating technique for obtaining multiparametric and asymptotically 
          Reissner-Nordstr\"om (RN) solutions.
\item A description of a class of regular  \bh\ solutions.
\end{enumerate}  
  Also, the authors consider only solutions with an electric field, which is important since, 
  in general, there is no electric-magnetic duality in NED. 

  Let us briefly discuss all these points. For better transparency, let us restrict ourselves to \ssph\
  metrics (an extension to the planar and pseudospherical symmetries is simple and evident)
  and apply more usual notations that those in \cite{GDC}. On the other hand, for completeness,
  systems with both electric and magnetic charges will be considered.
  
  It makes sense, for further discussion, to begin with reproducing some well-known facts on the
  \ssph\ Einstein-NED system according to \cite{Pel-T, k-NED1, k-NED2, dy2} (and many others)
  in the presence of both electric and magnetic charges. We consider the action
\beq            \label{S}
	S = \Half \int \sqrt {-g} d^4 x [R - L (f)], 	\qquad  f = F\mn F\MN, 
\eeq
  where $F\mn$ is the electromagnetic tensor, $L(f)$ is an arbitrary function, and we use 
  units in which $c = 8\pi G =1$. In the general \ssph\ metric 
\beq            \label{ds} 
	ds^2 = A(r) dt^2 - \frac{dr^2}{B(r)} - r^2 (d\theta^2+\sin^2 \theta d\phi^2),
\eeq
  the only possible nonzero components of $F\mn$ are $F_{tr} =- F_{rt}$ (a radial electric field) 
  and $F_{\theta\phi} = - F_{\phi\theta}$ (a radial magnetic field). The electromagnetic field
   equations $\nabla_\mu (L_f F\MN) = 0$ and $\nabla_\mu {}^*F\MN = 0$ (the asterisk 
   denotes duality) imply
\beq              \label{F_mn}
                      r^2 L_f F^{tr} = q_e, \cm    F_{\theta\phi} = q_m\sin\theta,
\eeq
  where $L_f \equiv dL/df$, and the constants $q_e$ and $q_m$ are the electric and magnetic 
  charges, respectively. The only nonzero SET components are 
\bearr           \label{SET}
            T^t_t = T^r_r = \frac L2 + f_e L_f, \quad T^\theta_\theta = T^\phi_\phi = \frac L2 - f_m L_f,
\yyy
	f_e = 2 F_{tr}F^{rt}=\frac{2q_e^2}{L_f^2 r^4}, \ \ \ 
	f_m = 2 F_{\theta\phi}F^{\theta\phi} = \frac {2q_m^2}{r^4},             \label{f_em}
\ear
   so that $f = f_m - f_e$. The equality $T^t_t = T^r_r $, through the Einstein equations, leads 
   without loss of  generality to $A(r) \equiv B(r)$, so that
\beq            \label{ds-A} 
	ds^2 = A(r) dt^2 - \frac{dr^2}{A(r)} - r^2 (d\theta^2+\sin^2 \theta d\phi^2),	
\eeq
  with only one unknown metric function $A(r)$. From the Einstein equation $G^t_t = - T^t_t$ 
  it follows
\beq          \label{A}
	A(r) = 1 -\frac{2M(r)}{r}, \qquad M(r) = \frac 12 \int T^t_t(r) r^2 dr,  
\eeq
  where $M(r)$ is called the mass function. This relation involves both electric and magnetic 
  charges according to \rf{F_mn} and \rf{SET} \cite{k-NED1, k-NED2}. A possible inclusion of 
  the cosmological constant $\Lambda$ adds the term $ -\Lambda r^2/3$ to the expression \rf{A}
  for $A(r)$, the corresponding solutions are discussed, e.g., in \cite{la1, la2, la3, la4}.

  Now we can pass on to discussing items 1--6.
  
\medskip\noi  
 {\bf 1. The Birkhoff theorem.} There was no need to prove this theorem anew for the system in 
  question because it is a special case of the extended Birkhoff theorem proved in \cite{bir1,bir2},
  where sufficient conditions for its validity were formulated as requirements to the SET 
  components, and these conditions ($T^r_t = 0$ and that some combination 
  $T^r_r + \const \cdot T^t_t $ should not depend on $A$ if, in the metric \rf{ds}, both $A$ and
  $B$ are allowed to depend on both $r$ and $t$) are manifestly fulfilled for the tensor \rf{SET}.
  Particularly, the Birkhoff theorem for systems with $T^r_r =T^t_t $ (``Dymnikova's vacuum'')
  was discussed in \cite{k-dym}.
  
  Another, more geometric formulation of the extended Birkhoff theorem was presented in 
  \cite{bir3, bir4}, and it includes, in particular, systems with SETs of Segre types      
  [(111,1)] and [(11) (1,1)] that correspond to a cosmological constant and NED, respectively.
  
   One should note that the meaning of the (extended) Birkhoff theorem is not that a certain 
   kind of matter in space-times of given (here, spherical) symmetry necessarily creates 
   a static metric, but that this metric, due to the field equations, necessarily contains an 
   {\it additional\/} symmetry, not initially postulated. The corresponding Killing vector may be
   timelike (then the metric is static), spacelike (as happens in time-dependent \bh\ interiors)
   or null (see examples, e.g., in \cite{bir5}). This important circumstance was not mentioned in 
   \cite{GDC}.    
   
\medskip\noi  
 {\bf 2. Algebraic types of $T\mN$ and $G\mN$.}  As already said, having the structure
  \rf{SET}, the SET of NED with spherical symmetry belongs to the Segre type  [(11) (1,1)]
  that corresponds to two different pairs of eigenvalues. This is certainly well known and 
  cannot be regarded a result. However, of certain interest is the observation that in the
  Einstein-NED system the traceless Ricci tensor $S\mN := R\mN - \frac 14 \delta\mN R$ has the
  form $S\mN =  S\,\diag(1, -1, 1, -1)$.  
  
  What the authors of \cite{GDC} call a {\it theorem} (top of the right column on page 4),
  sounds really strange: ``Besides the vacuum with $\Lambda$ solutions, static Schwarzschild-like
  metrics only allow electromagnetic solutions to the Einstein (linear or nonlinear) 
  electrodynamics equations.''  First, the authors repeatedly use the  words ``Schwarzschild-like 
  metrics'' but {\it nowhere\/} define what they mean by them. Even if this term is used somewhere
  else, it is not widely known and must be clearly defined. Second, very probably 
  ``Schwarzschild-like'' means the metric \rf{ds-A} with any $A(r)$. But anyway, the algebraic type 
  of the SET certainly  does not uniquely prescribe the kind of matter it belongs to. 
  For example, the Segre type [(11) (1,1)] of SET pertains not only to NED but also to non-Abelian 
  Yang-Mills fields.
   
\medskip\noi     
{\bf 3. The inverse integration method.} For electric solutions ($q_e \ne 0,\ q_m=0$), the inverse
  integration method is formulated in \cite{GDC}, presenting $F_{rt}$ and $L(f)$ in terms of the 
  ``structure function'' $Q(r) = A(r)/r^2$ in \eqs (18) and (19). In our notations, in terms of the 
  metric function $A(r)$, we have equivalently
\bearr                             \label {Fe} 
	2q_e F_{tr}(r) = -1 + A - \Half r^2 A'',
\\ \lal                               \label{Le}
	L(r) = - \Lambda - A'' - 2\frac{A'}{r}	
\ear
  (the prime stands for $d/dr$). This gives the quantities $F_{tr}$ (hence $f = - f_e = - 2 F_{tr}^2$) 
  and $L$ as functions of $r$.  It is, however, important but ignored in \cite{GDC}, that 
  with chosen $A(r)$ (or $Q(r)$),  the function $f(r)$ will not always be monotonic, and thus
  it is not always possible to obtain an unambiguous Lagrangian function $L(f)$. 
  
  Unlike that, for systems with pure magnetic charge ($q_e = 0,\ q_m \ne 0$), the function 
  $f(r) = f_m = 2q_m^2/r^4$ is monotonic, hence for given $A(r)$ we always obtain a well-defined 
  Lagrangian function $L(f)$ \cite{k-NED1}: as follows from \rf{A} with possible $\Lambda \ne 0$,
\beq                                  \label{Lm}
		L(r) = \frac{4M'}{r^2} = -2\Lambda + \frac{2}{r^2}(1 - A - rA'). 
\eeq    
    
  In the dyonic case with both nonzero $q_e$ and $q_m$, the situation is more complicated
  \cite{dy1, dy2, dy3}: given $A(r)$, there is no direct expression for $L(r)$; instead of \rf{Lm}, 
  we obtain from \rf{A} 
\beq                                  \label{L_dy}
		 \frac{4M'}{r^2} = -2\Lambda + \frac{2}{r^2}(1 - A - rA') = L + \frac{4q_e^2}{L_f r^4}.
\eeq
     
\medskip\noi      
{\bf 4. Linear superpositions of solutions.} As noticed in \cite{GDC}, \eqs (18) and (19) 
  (equivalent to \rf{Fe} and \rf{Le} in the present Comment) are linear in the structure function
  $Q(y)$ or, equivalently, in the metric function $A(r)$ in the present notations. It then directly
  follows that if each of the functions $Q_i(y)$ (or $A_i(r)$) describes a solution to the 
  NED-Einstein equations with electric charges $q_{ei}$ and the quantities $q_{ei} F_{tr}(r)$,
  $L_i(r)$ and $\Lambda_i$, then their linear combinations (with constant coefficients $c_i$) also 
  describe electric solutions, in which the quantities $q_e F_{tr}(r)$, $L(r)$ and $\Lambda$ are 
  linear combinations of the constituent quantities with the same coefficients.
  
  The authors have formulated this result as a theorem: ``For static Schwarzschild metrics coupled 
  to electrodynamics (linear and nonlinear) and a $\Lambda$ term (if any), any linear superposition 
  of structural functions leads to linear superpositions of Lagrangian functions and the
  corresponding electromagnetic field functions.'' 
  
  This formulation, as well as the unnumbered equation after it, are not precise, even forgetting
  that the term ``Schwarzschild metric'' is used here in an unusual manner. The following points are 
  missing:  (i) For $\sum_i c_i A_i(r)$ to satisfy \rf{Fe}, it is required  $\sum_i c_i =1$; (ii) not $F_{tr}$ 
  but $q_e F_{tr}$ is a subject of superposition (the statement on $F_{tr}$ in [1] is correct only if 
  the charge $q_e$ is the same in all constituent solutions), and (iii) the resulting $\Lambda$ is 
  $\sum_i c_i\Lambda_i$.
  
  One should add that, as before, in all thus obtained electric solutions one should take care of the 
  monotonicity of  $f(r) = -2 F_{tr}^2$, otherwise $L(f)$ is ill-defined.  
  
  In the magnetic case, \eq \rf{Lm} is also linear with respect to $A(r)$, therefore emerges a similar 
  superposition method of constructing new solutions from known ones, but the magnetic charge
  values are not directly related to \rf{Lm}, and this issue should be analyzed separately.  
  
\medskip\noi  
{\bf 5. Multiparametric and asymptotically RN solutions.} The above-described superpositions
   are characterized in \cite{GDC} as a generating technique for obtaining multiparametric 
   solutions. Indeed, it allows for obtaining new solutions from known ones. Though, concerning 
   the number of parameters, let us recall that actually, if $L(f)$ is not specified, we have an 
   arbitrary function $A(r)$ (or $Q(y)$ in \cite{GDC}), which can be endowed with {\it any\/}
   number of parameters. As to solutions with RN asymptotic behavior,  it is clear that if $A(r)$
   is specified with a proper large $r$ behavior, the whole solution will also behave properly at large $r$.
   The example of a multiparametric electric solution discussed in \cite{GDC} confirms that. 
   Similar examples of magnetic solutions can also be constructed.    

\medskip\noi  
{\bf 6. Regular black holes and solitons.}
  It is correctly said \cite{GDC} that regularity at the center $r=0$ (as at any other location) requires 
  finiteness of all curvature invariants. However, there is no need to calculate them for each 
  particular solution since it is well known that the metric \rf{ds-A} is regular at $r=0$ if and 
  only if $A(r) = 1 + \const \cdot r^2 + o(r^2)$ as $r\to 0$ (see, e.g., \cite{br-book}). 
  Any such function provides a regular center, both in the electric and magnetic cases. 
  Moreover, a superposition of solutions regular at $r=0$ is also regular at $r=0$ under 
  the evident condition $\sum_i c_i =1$.
     
  However, a well-known important property of electric solutions, not mentioned in \cite{GDC},
  is that a Lagrangian function $L(f)$ providing a solution with a regular center cannot have 
  a Maxwell weak field limit ($L\sim f$ as $f\to 0$) \cite{B-Shi}. At such a center the electric field 
  should be zero, so that $f \to 0$, but it then follows from the field equations that $L_f \to \infty$ 
  as $r\to 0$, which means a strongly non-Maxwell behavior at small $f$. In dyonic solutions 
  ($q_e\ne 0,\ q_m \ne 0$), a regular center also requires a 
  non-Maxwell weak field limit of $L(f)$ \cite{k-NED1}. Only pure magnetic solutions 
  are compatible with a correct weak-field limit of $L(f)$: in this case, at a regular center, 
  $f = 2q_m^2/r^4 \to \infty$ but $L(f) \to \const < \infty$, providing finite limits of both the SET 
  components and the curvature invariants.  
  
  An electric solution can have a regular center (it describes a black hole if there are zeros of $A(r)$ 
  at $r>0$ or a solitonic particlelike object if $A(r) >0$ at all $r$), where $L_f \to \infty$ as $f\to 0$,
  and a RN asymptotic behavior at large $r$, where $L \propto f$ at small $f$, which, however, means 
  that these are {\it different\/} functions $L(f)$. In other words, different NED theories are acting in 
  different parts of space. It is an example of what happens if, in a particular solution, $f(r)$ in not 
  monotonic. Pure magnetic solutions are free from this shortcoming. A more detailed discussion of 
  such situations can be found in \cite{k-NED1, dy2, k-comment}.
  
  To conclude, the paper \cite{GDC}, containing some interesting results and observations,  
  is not free from significant gaps and inaccuracies which I tried to fill or correct in this Comment.

\subsection*{Acknowledgments}

  I thank Milena Skvortsova and Sergei Bolokhov for helpful discussions. 
  The work was partly performed within the framework of the Center FRPP 
  supported by MEPhI Academic Excellence Project (contract No. 02.a03.21.0005, 27.08.2013).
  The work was also funded by the RUDN University Program 5-100.


\end{document}